# Layer Number Determination and Thickness-dependent Properties of Graphene Grown on SiC

Wenjuan Zhu, Christos Dimitrakopoulos, Marcus Freitag and Phaedon Avouris

*Abstract*— The electronic properties of few-layer graphene grown on the carbon-face of silicon carbide (SiC) are found to be strongly dependent on the number of layers. The carrier mobility is larger in thicker graphene because substrate-related scattering is reduced in the higher layers. The carrier density dependence of the mobility is qualitatively different in thin and thick graphene, with the transition occurring at about 2 layers. The mobility increases with carrier density in thick graphene, similar to multi-layer graphene exfoliated from natural graphite, suggesting that the individual layers are still electrically coupled in spite of reports recording non-Bernal stacking order in C-face grown graphene. The Hall coefficient peak value is reduced in thick graphene due to the increased density of states. A reliable and rapid characterization tool for the layer number is therefore highly desirable. To date, AFM height determination and Raman scattering are typically used since the optical contrast of graphene on SiC is weak. However, both methods suffer from low throughput. We show that the scanning electron microscopy (SEM) contrast can give similar results with much higher throughput.

## I. INTRODUCTION

Graphene is a two dimensional atomic layer of carbon atoms forming a honeycomb crystal lattice [1]. The high intrinsic mobility in graphene [2-3] makes it an attractive material for high speed electronics, especially RF circuits. Thin films of graphene can be formed by exfoliation of bulk graphite [4], by CVD growth on certain metals [5-6], or by epitaxial growth on SiC by high temperature decomposition of its surface and sublimation of Si [7-12]. Graphene grown on SiC has the advantages of uniform coverage and a coherent structure at wafer-scale. None of these advantages are feasible – at least up to date – in the case of exfoliated graphene, and structural coherence is lost in the polycrystalline graphene CVD films at wafer scale. Furthermore, graphene grown on semi-insulating SiC does not have to be transferred to another insulating substrate, as is the case with CVD grown graphene on metals. Therefore, graphene grown on SiC has recently become the subject of intense research.

For exfoliated graphene, the number of layers can be reliably determined by the light reflectance method, based on the optical contrast between graphene and the underlying $Si/SiO_2$ substrate [13-15]. The method relies on interference enhancement due to the oxide layer, which needs to be close to 90nm, 280nm, etc in height. In epitaxial graphene grown on SiC, graphene is directly sitting on SiC without any oxide, and the contrast is very poor (nearly invisible to the eye). Low-energy electron microscopy (LEEM) [16] can determine the graphene layer number, but the sample size and the field of view are limited and the measurements are time consuming. In this work, we report that the SEM contrast can provide information on the number of layer accurately, and the entire wafer can be mapped quickly. We correlate the SEM contrast with AFM height measurements and the Raman G-band position, showing consistency across all three methods. The electronic properties (Hall mobility and Hall coefficient) are measured for different thicknesses, and we find important qualitative and quantitative differences in thin and thick graphene.

## II. EXPERIMENT:

We grew graphene on the C-terminated face of quarter wafer pieces of high purity semi-insulating (HPSI) 4H($000\bar{1}$) SiC wafers (2 inch diameter) that had a chemically-mechanically polished (CMP) epitaxy-ready surface on their C polar face. Graphene growth took place in a UHV chamber (base pressure of ~ $3 \times 10^{-10}$ Torr) equipped with a custom-designed, inductively heated hot-zone comprising a cylindrical graphite susceptor. The SiC pieces were loaded on a graphite carrier and then loaded via a load-lock into the hot zone area. After degassing at 810 °C for 21 min in vacuum (P<$1 \times 10^{-8}$ Torr), the SiC was cleaned mainly from oxide contamination by annealing at 810 °C under disilane flow (20% disilane in He) for 10 min. After this cleaning step, the SiC wafer was annealed at 1450 °C for 2 min under Ar flow at a pressure of $3.2 \times 10^{-4}$ Torr and then was allowed to cool down in Ar. Subsequently, metal alignment marks were formed by lift-off and graphene Hall-bar structures were fabricated by photoresist patterning and $O_2$ plasma etching. Following that, SEM, AFM and Raman characterization was performed, and then source/drain and sensing terminals were formed using Ti/Pd/Au metallization and lift-off. The SEM measurements were taken at 3KV. The AFM images were taken in tapping mode. Silicon nitride gate

Manuscript submitted on Sep. 23, 2010. This work is supported by DARPA under contract FA8650-08-C-7838 through the CERA program.

The authors are with IBM Thomas J. Watson Research Center, Yorktown Heights, NY 10598, USA. Corresponding authors are Wenjuan Zhu (914-945-2894; wenjuan@us.ibm.com) and Phaedon Avouris (914-945-2722; avouris@us.ibm.com).

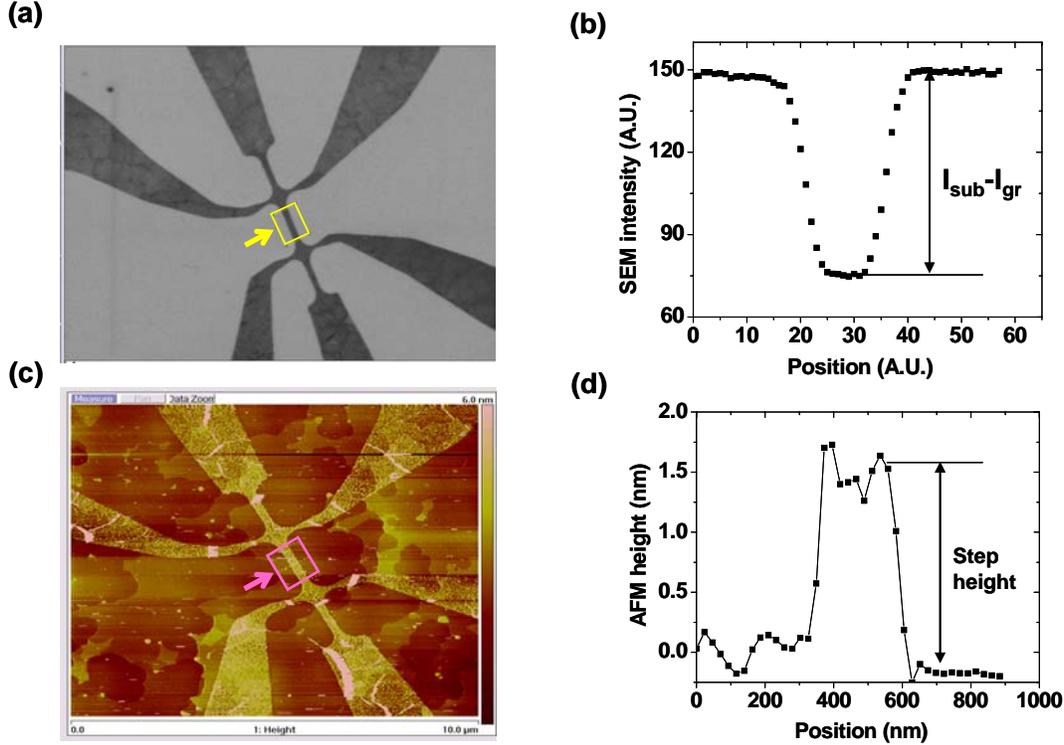

**Fig. 1.** SEM (**a**) and AFM (**c**) images of graphene grown on C-face SiC. (**b**) and (**d**) shows the profile of the SEM intensity and AFM height across the channel. Here the SEM intensity and the AFM height are the average of signal along the channel direction in the rectangle region as marked in the images.

dielectric was deposited by PECVD at 400°C [17]. Top gate electrodes were then formed using Ti/Pd/Au. Hall mobility and Hall coefficient were measured using magnetic field of +/- 2Tesla at temperature of 300K and 4.2K.

III. RESULTS:

Figure 1(a) shows an SEM image of a graphene Hall bar on the C-face of SiC. The dark region is graphene and the bright region is the exposed SiC substrate after the graphene has been etched away using $O_2$ plasma. The profile of SEM intensity across the Hall bar channel is shown in Fig. 1(b). The SEM intensity in the graphene channel is lower than the intensity on the SiC substrate because graphene is more conductive than SiC substrate and fewer secondary electrons are reflected back to the secondary electron detector. Figure 1(c) shows an AFM image of the same Hall bar and Fig. 1(d) shows the corresponding height profile across the Hall bar. For this Hall bar, the graphene channel appears about 1.5nm higher than the substrate. Figure 2 shows the AFM height vs. SEM contrast for 26 Hall bars. Here the AFM height is defined as the height difference between the channel and the nearby substrate, and the SEM contrast is defined as $(I_{sub} - I_{gr})/I_{sub}$, where $I_{gr}$ is the SEM intensity in the graphene channel and $I_{sub}$ is the SEM intensity on the nearby substrate. There is a strong correlation between AFM height (and thus layer number) and SEM contrast. The thicker the graphene, the higher the SEM contrast due to the increased conductivity of the graphene layer.

According to the "C-corrugated" model in [10], the distance between the first graphene layer and the silicon plane in the interface layer is 0.325nm for graphene on C-face of SiC. The distance between graphene layers in graphite was reported as 0.335nm [18]. Assuming the interfacial layer composition and geometry follows the "C-corrugated" model and the distance between subsequent graphene layers are similar to those in graphite, we can estimate the number of graphene layers based on the AFM height, as marked in Figure 2. Based on the correlation between the AFM height and SEM contrast, we can establish the correlation between SEM contrast and the number of graphene layers, as marked in Figure 2.

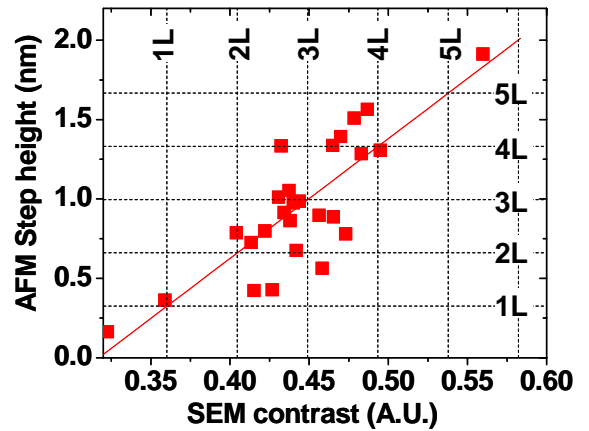

**Fig. 2.** AFM height vs. SEM contrast for 26 graphene Hall-bars.

Raman spectroscopy was additionally used to characterize the Hall bars. Figure 3(a) shows Raman spectra for graphene Hall bars with SEM contrast ranging from 0.36 to 0.56, which corresponds to about 1 to 6 graphene layers. Both G and 2D bands were fitted with Lorentzians and the peak area, width, and position determined. The 2D band area and the G band position are most affected by the layer number. We found that the G-prime area increases with layer number, while the G band energy decreases. In Fig. 3(b) we plot the Raman G band position vs. SEM contrast (and thus layer number). The G band position decreases by 8cm$^{-1}$ between 1 and 6 layers. In graphene grown on SiC, the first layer from the SiC interface is usually heavily doped by the substrate [19-20]. As the graphene gets thicker, the top layer is further away from the SiC interface, and thus a larger portion of the channel is less doped. Therefore, the energy of the G-band decreases, consistent with studies of doping in exfoliated graphene [21]. From the correlation between SEM contrast and Raman G-band position, we establish a correlation between Raman G-band shift and graphene layer number, as marked in Figure 3. Note that at layer 6, the Raman G-band position has reached 1584cm$^{-1}$, which is generally considered to be the value for undoped graphene.

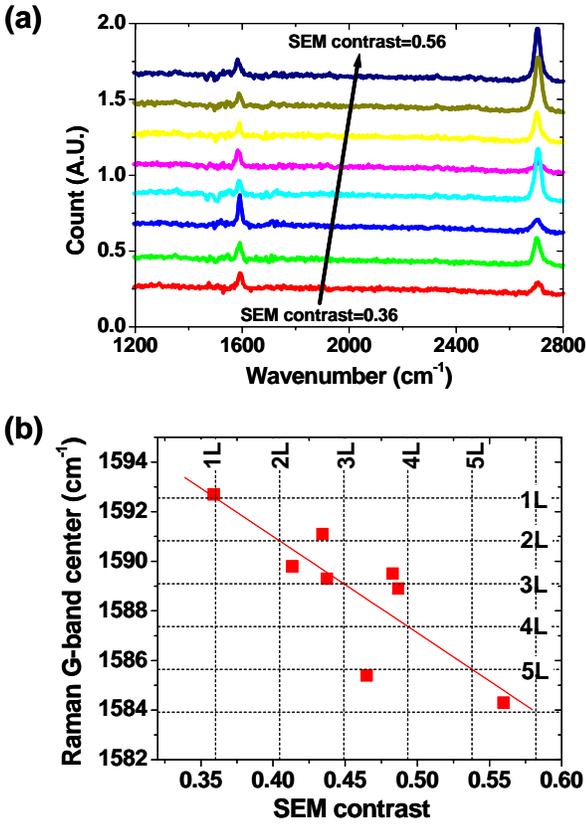

**Fig. 3.** (a) Raman spectra of graphene Hallbars on SiC with SEM contrast from 0.36 to 0.56. (b) Raman G-band position vs. SEM contrast for the above graphene Hall-bars.

The number of layers of graphene also significantly influences the electrical properties. Figure 4 shows the Hall coefficient vs top gate voltage for Hall bars with SEM contrast from 0.36 to 0.49 (about 1 to 4 layers) measured at 4.2K and 300K. The Hall coefficient is defined as $R_H = V_H / I_H B$, where $V_H$ is the measured Hall voltage, $I_H$ is the constant source current and $B$ is the applied magnetic field. Figure 4 (b) shows the Hall coefficient peak height versus SEM contrast. As the graphene layer number increases, the Hall coefficient peak is reduced significantly. This can be explained by the following considerations. Near the Dirac/neutrality point, electron and hole puddles can form due to variations of the surface electrostatic potential [22-23]. If we assume that the area of the hole and electron puddles is equal in size and simplify the spatial electrostatic potential to a step function with the peak to peak height of $\pm\Delta$, the electron and hole carrier densities can be expressed by the following equations:

$$n_e(E_F) = \int_{-\Delta}^{\infty} \frac{1}{2} D(E+\Delta) f(E) dE + \int_{\Delta}^{\infty} \frac{1}{2} D(E-\Delta) f(E) dE \quad (1)$$

$$n_h(E_F) = \int_{-\infty}^{-\Delta} \frac{1}{2} D(-E-\Delta)[1-f(E)] dE + \int_{-\infty}^{\Delta} \frac{1}{2} D(-E+\Delta)[1-f(E)] dE \quad (2)$$

where $E_F$ is Fermi level and $f(E)$ is the Fermi-Dirac distribution function. The density-of-states in single-layer graphene is $D_{SL}(E) = 2E/[\pi(\hbar v_F)^2]$ and the one in multi-layer graphene with coupled layers is $D_{ML\_coupled}(E) = 2m/(\pi\hbar^2)$, where m is the effective mass of the graphene [24]. The ambipolar Hall coefficient is given by [25]: $R_H = \dfrac{n_h \mu_h^2 - n_e \mu_e^2}{e(n_h \mu_h + n_e \mu_e)^2}$. Assuming that the electron and hole mobility are similar, this equation can be simplified to

$$R_H = \frac{n_h - n_e}{e(n_h + n_e)^2}. \quad (3)$$

Based on equations (1-3), we can calculate the Hall coefficient as a function of the Fermi energy and extract the Hall coefficient peak height. Figure 4 (c) shows the calculated Hall coefficient peak height as a function of graphene layer numbers with $\Delta = 80 meV$ and $\Delta = 100 meV$ at 4.2K and 300K. As the graphene layer number increases, the effective mass and density of states increases, resulting in lower Hall coefficient peak values. As the temperature increases, the Hall coefficient peak value also decreases, due to the thermal broadening in Fermi-Dirac distribution. These trends were indeed observed in the measurement shown in Figure 4(b). Note that the Hall coefficient peak value is also influenced by the variation in the electrostatic potential. The larger the electrostatic potential, the lower the peak height, due to the larger carrier density induced by the additional electrostatic charges. Moreover, even in uncoupled layers, the density of states of multi-layer graphene will still increase with increasing number of layers, which would result in a reduced Hall coefficient peak height.

magnetic field, and $e$ is the electron charge. We can see that as the graphene layer number increases, the carrier density dependence of the Hall mobility changes. For thin graphene (~1 layer), the mobility decreases with increasing carrier density, while for thick graphene (~4 layers), the mobility increases with increasing carrier density. For medium thickness graphene (~2 layers), the carrier density dependence is in between these above two cases, i.e. nearly independent of carrier density. This can be explained by the difference of the density of states in single-layer graphene and multi-layer graphene. At low temperatures, the dominant scattering mechanism is Coulomb scattering by impurities and short-range scattering by defects. The overall mobility can be found using a Matthiessen's rule $\mu_{total}^{-1} \approx \mu_{sr}^{-1} + \mu_{C}^{-1}$, where $\mu_{sr}$ is the mobility limited by short-range scattering and $\mu_{C}$ is the mobility limited by Coulomb scattering. In single-layer graphene, the density of states is proportional to Fermi energy: $D_{SL}(E) \propto E$, which will result in $\mu_{sr} \propto 1/n$ and $\mu_{C}$ is constant, thus the overall mobility at low temperature decreases with increasing carrier density [26]. For multi-layer graphene however, the density of states is constant, which results in a constant $\mu_{sr}$ and $\mu_{C} \propto n$, thus the overall mobility at low temperature increases with increasing carrier density [27]. When there is a mixture of mono- and multi-layers, the carrier density dependence will be in between these two extremes, i.e. nearly independent of carrier density.

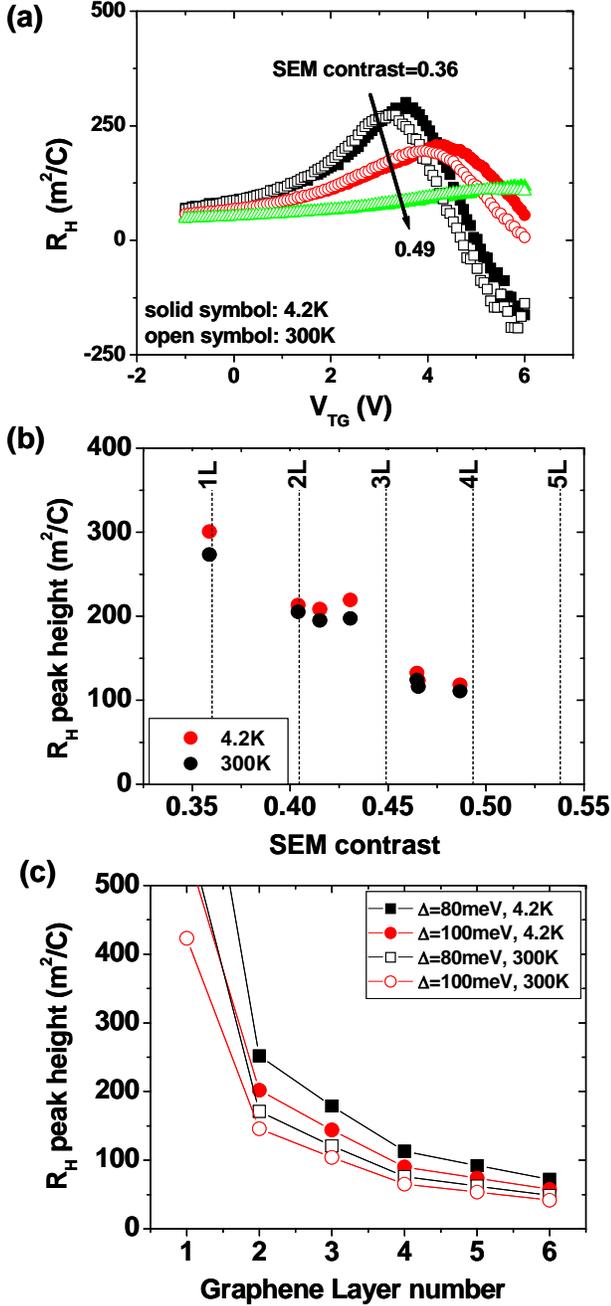

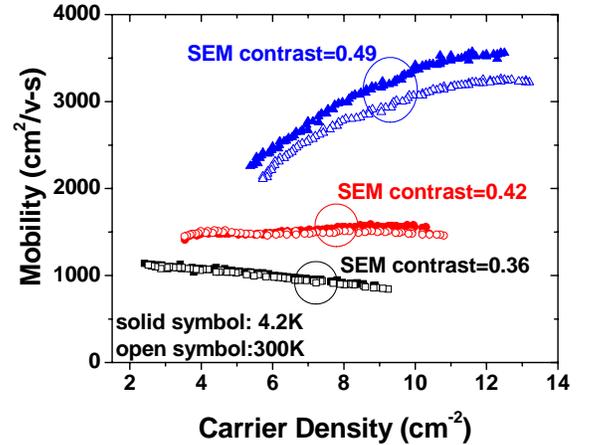

**Fig. 4.** (a) Hall coefficient of graphene Hall-bars with SEM contrast from 0.36 to 0.49 measured at 4.2K and 300K. (b) Measured Hall coefficient peak height vs. SEM contrast. (c) Calculated Hall coefficient peak height vs. layer number for graphenes with Δ=80meV and Δ=100meV at 4.2K and 300K.

**Fig. 5.** Hall mobility as a function of carrier density for graphene Hall-bar devices with SEM contrast of 0.49, 0.42 and 0.36 measured at 4.2K and 300K.

Finally, the mobility of the graphene channel is also influenced by the graphene layer number. Figure 5 shows the Hall mobility as a function of carrier density for graphene Hall bars with SEM contrast of 0.36, 0.42 and 0.49, which corresponds to about 1 layer, 2 layers and 4 layers respectively. The carrier density n was extracted from the Hall voltage $V_H$: $n = I_H B / e|V_H|$, where $I_H$ is the current, $B$ is the magnetic field, and $e$ is the electron charge.

The mobility increase with carrier density in multi-layer graphene indicates the presence of layer to layer coupling in graphene grown on SiC, and possibly a parabolic band structure just like in exfoliated few-layer graphene. This would be consistent with band structures measured by ARPES (Angle-Resolved Photoemission Spectroscopy) for 1 to 4 layers of graphene grown on Si-face SiC [28]. For thicker graphene grown on the C-face of SiC (11 layers), however, it was reported that the ARPES showed linear band structure

[29], possibly due to the increased sensitivity of ARPES to the top 3–4 surface layers. Most likely, however, there is a coexistence of coupled and uncoupled layers and variable layer stacking which is dependent on the condition of the synthesis.

Importantly, as the graphene layer number increases, the mobility increases as well (for example at carrier density of $8.5 \times 10^{12} cm^{-2}$, the mobility increases from ~900 $cm^2$/v-s for 1 layer to ~3100 $cm^2$/v-s for 4 layers in our graphene Hall bars). This should be due to a reduced charged impurity scattering from the substrate, as the top layers got further away from the substrate in thicker graphene. As the temperature increases, the mobility decreases slightly, due to the increase of scattering from the gate dielectric surface optical phonons and graphene phonons [30].

## IV. SUMMARY

In summary, we found that there is strong correlation between SEM contrast, AFM height, Raman G-band position, Hall coefficient peak height and Hall mobilities for graphene grown on C-face SiC. As the number of graphene layers increases, the SEM contrast increases due to the increased conductivity in the graphene channel, the AFM height increases due to the added graphene layers, and the position of the Raman G-band decreases due to the reduced doping in the top graphene layers. Furthermore, as the number of graphene layers increases, the Hall coefficient peak height decreases due to the increased density of states. The carrier density dependence of Hall mobility changes from single-layer-like (mobility decreases with increasing carrier density) to multi-layer-like (mobility increases with increasing carrier density), similar to the case of exfoliated graphene. This indicates that there is strong coupling between layers in graphene grown on the SiC.

## V. ACKNOWLEDGEMENT

We would like to thank B. Ek, J. Bucchignano, and G. P. Wright for their contributions to device fabrication. We are also grateful to Y.-M. Lin, D. Farmer, F. Xia, H.-Y. Chiu, Y. Wu, and C.Y. Sung for their insightful discussions.